\documentclass{article}
\usepackage[ansinew]{inputenc}
\usepackage{latexsym}
\input{epsf}
\newcommand\bearst{\begin{eqnarray*}}
\newcommand\eearst{\end{eqnarray*}}

\newcommand{\slpe}{\raise.15ex\hbox{$/$}\kern-.57em\hbox{$p$}}
\newcommand{\slpartial}{\raise.15ex\hbox{$/$}\kern-.57em\hbox{$\partial$}}
\newcommand{\slp}{\raise.15ex\hbox{$/$}\kern-.57em\hbox{$p$}}
\newcommand{\slq}{\raise.15ex\hbox{$/$}\kern-.57em\hbox{$q$}}
\newcommand{\slk}{\raise.15ex\hbox{$/$}\kern-.57em\hbox{$k$}}
\newcommand{\sla}{\raise.15ex\hbox{$/$}\kern-.57em\hbox{$a$}}
\newcommand{\slA}{\raise.15ex\hbox{$/$}\kern-.57em\hbox{$A$}}
\newcommand{\slB}{\raise.15ex\hbox{$/$}\kern-.57em\hbox{$B$}}
\newcommand{\slD}{\raise.15ex\hbox{$/$}\kern-.57em\hbox{$D$}}
\newcommand{\slb}{\raise.15ex\hbox{$/$}\kern-.57em\hbox{$b$}}
\newcommand{\slc}{\raise.15ex\hbox{$/$}\kern-.57em\hbox{$c$}}
\newcommand{\sld}{\raise.15ex\hbox{$/$}\kern-.57em\hbox{$d$}}
\newcommand{\slW}{\raise.15ex\hbox{$/$}\kern-.57em\hbox{$W$}}
\newcommand{\slP}{\raise.15ex\hbox{$/$}\kern-.57em\hbox{$P$}}

\newcommand{\be}{\begin{equation}}
\newcommand{\ee}{\end{equation}}
\newcommand{\bear}{\begin{eqnarray}}
\newcommand{\ear}{\end{eqnarray}}
\newcommand{\ba}{\begin{eqnarray*}}
\newcommand{\ea}{\end{eqnarray*}}
\renewcommand{\theequation}{\arabic{section}.\arabic{equation}}

\def\qM{q^{(+)}}
\def\qm{q^{(-)}}
\def\piM{\Pi_0^{(+)}}
\def\pim{\Pi_0^{(-)}}
\begin{document}

\title{Thermofield-Bosonization on Compact Space }
\author{R. L. P. G. Amaral and L. V. Belvedere \\
\it{ Instituto de F\'{\i}sica}\\
\it{Universidade Federal Fluminense}\\
\it{Av. Litor\^anea S/N, Boa Viagem, Niter\'oi, CEP. 24210-340}\\
\it{Rio de Janeiro - Brasil}}

\date{\today}

\maketitle

\begin{abstract}

We  develop the construction of fermionic fields in terms of bosonic ones  to describe free and
interaction models in the circle, using thermofielddynamics. The description in the case of finite temperature is developed for both normal modes and zero modes. The treatment extends the thermofield-bosonization for periodic space.

\end{abstract}

\section{Introduction}

We have discussed quite recently the symmetry aspects of models which describe charge-density waves in incommensurate case within a field theoretic
set-up, \cite{cdw}. The bosonization of the fermions, \cite{AAR}, has shown to be a most valuable tool. The field theoretic models that describe the phenomenon are constructed with fields in an infinite dimensional space, based upon lower dimensional versions of models used in the description of relativistic elementary particles. In the commensurate case, however, the periodicity of the net of charge distribution and of the sound waves become important. Also the finite  extent of the sample needs to be addressed. The extension of the field theoretical models which encodes periodic boundary conditions has to be considered in this case. 
The diversity of new physical phenomena which are described by lower dimensional fermionic field theory models, as for example in graphene and topological insulators, constitute further motivations for this study. 

A description of the bosonization of spatially periodic fields has been presented, within a condensed matter context, a long time ago by
Haldane, \cite{haldane}, and has been the subject of much renewed interest recently\cite{recentperiodic1,recentperiodic2,recentperiodic3}. However,  an approach which, based on field theoretic methods, stresses the parallelism with the infinite space case is lacking.
 We consider here the quantization of scalar and
fermionic free fields and the bosonization of fermionic fields in 1+1 dimensions for spatially periodic boundary conditions. The purpose is to stablish which are the essential ingredients that have to be considered in order to account for the space periodicity, what seems to be obscured when the peculiarities of the additional demands of the condensed matter context are taken into account. 

A further kind of periodicity emerges, however, due to the consideration of a heat bath, as required by the KMS\cite{KMS, Haag} conditions: the periodicity in the imaginary time. The simultaneous occurrence of  both kinds of periodicities are considered here. In order to keep a close parallelism with the infinite space case we employ thermofielddynamics methods. These methods have been considered quite recently by us\cite{ABR, ABR2}, in the bosonization of infinite space fields at finite temperature and have lead to a much transparent description, as has been witnessed in \cite{qedt}, which emphasizes the similarities with the zero temperature case. This work generalizes the thermofieldbosonization procedures to the case of concurrence of both kind of periodicities and with this aims differs from the contributions to the bosonization of compact space fields reported in \cite{qedc}

The paper is organized as follows: We begin in Section 2 by considering the construction of free fields in compact spaces, with periodic boundary conditions. This streamlines the notation and makes the paper self contained. In section 3 we describe the bosonization of the periodic fields at zero temperature. The comparison with the bosonization of fields in infinite space and finite temperature will be stressed. In section 4 we add temperature and present compact expressions for the fields and their 2-point functions within thermofielddynamics. In section 5 the bosonization is discussed in the case of finite temperature and a formalism to encode the contribution of the heat bath is developed.  After discussing the the results in the section 6 and presenting some technical details in the first two appendices,  in the appendix C we discuss the completeness of the operator bosonization by computing the four point fermionic functions using the bosonization prescription.

\section{ Free Fields With Periodic Conditions }


We will adopt the strategy of building models of fields with periodic boundary by imposing the minimal requirements besides the ones borrowed from relativistic quantum field theory.  Let us start by presenting the quantization of the hermitian scalar field in two space-time dimensions with periodic boundary conditions.
 
 In terms of light-cone variables, $x^\pm=x^0\pm x^1$ ( and $\partial_\pm=\partial_0\pm \partial_1$) the scalar field obeys the equation of motion $
 \partial_+\partial_-\phi =0$,  so that $\phi(x)=\phi(x^+)+\phi(x^-)$. 
Imposing spatially periodic boundary conditions, $\phi(x^\pm)=\phi(x^\pm+L)$, the mode expansion turns out to be given by

\be \label{me}
\phi(x^\pm)=\sum_{k=1}^\infty \frac{1}{2\sqrt{\pi k}}\left[ e^{-i\frac{2\pi k}{L}x^\pm}e^{-\epsilon k} a_k^{\pm}+e^{i\frac{2\pi k}{L}x^\pm} e^{-\epsilon k} a_k^{\pm\dagger}\right],
\ee 
where we have inserted the ultraviolet convergence factor $\epsilon$.

In order to obtain the equal-time canonical commutation relations, $
\left[\phi(x),\dot{\phi}(x^{\prime})\right]_{x^1\rightarrow x^{\prime 1}}\approx i\delta(x^1-x^{\prime 1})$,
we require that $\left[a_k^{\pm},a_{k^\prime}^{\pm\dagger}\right]=\delta_{k^\prime ,k}$. 
This results, however, in

\be\label{commutationrel}
\left[\phi(x^\pm),\dot{\phi}(x^{\prime\pm})\right]=\frac i{2L}\lim_{\epsilon\rightarrow 0}\sum_{k=1}^\infty\left[e^{-\left(\frac{i2\pi \Delta x^\pm}{L}+2\epsilon\right)k}+e^{\left(\frac{i2\pi \Delta  x^\pm}{L}-2\epsilon\right)k}\right]=\frac i2 \delta_L(\Delta x^\pm)-\frac{i}{2L}\ee
where $\delta_L$ represents the delta function with period $L$, $\delta_L(x+L)=\delta_L(x)$, and $\Delta x=x-x^\prime$. The constant term of the last right side term, which spoils the strict canonical commutation relations, can be
evaded by adding proper zero modes terms in the eq. (\ref{me}). 

From the mode expansion (\ref{me}) we obtain for the two-point function 

\be \label{discrete1}
D(\Delta x^\pm)\equiv <0|\phi(x^\pm)\phi(x^{\prime\pm})|0>=\sum_{k=1}^\infty \frac 1{4\pi k}e^{-\left(\frac{i2\pi \Delta x^\pm}{L}+2\epsilon\right)k}
=\frac {-1}{4\pi}\ln\left(1-e^{-2\epsilon-\frac{i2\pi \Delta x^\pm}{L}}\right),
\ee
which is the main result of this sub-section. The final expressions are periodic in space, and since they are functions of the light-cone variables, $x^\pm$, they are also periodic in time. This is enforced by the $\epsilon$-regulator, which avoids the crossing of the branch-cut line of the logarithm.

For later reference, it is instructive to obtain the Fourier representation of the two-point function. As shown in the  Appendix A

\bear
D(\Delta x^\pm) &=& \frac{i\pi}{4L}( \Delta x^\pm -\frac L2)+\ln 2+
\int_0^\infty \frac {dp}{4\pi p}\left[\left( e^{-ip(\Delta x^\pm-i\epsilon)}\frac 1{1-e^{-ipL}}+e^{ip(\Delta x^\pm+i\epsilon)}\frac{e^{-ipL}}{1-e^{-ipL}}\right)-\right.\nonumber\\
&&\left.\left. e^{-ip(L/2-i\epsilon)}\frac 1{1-e^{-ipL}}+e^{ip(L/2+i\epsilon)}\frac{e^{-ipL}}{1-e^{-ipL}}\right)\right].
\ear

 The first term can be reproduced from zero mode contributions. The last term, apart from an additive renormalization, can be obtained from field theory treatment, within thermofielddynamics, of the infinite space scalar field, as we will discuss elsewhere.

 \subsection{ Spatially Periodic Dirac Field}
 
Let us construct a solution do the Free Dirac field with anti-periodic boundary conditions in space, $\Psi(x+L,t)=-\Psi(x,t)$.
The mode expansion is given by

\be\label{diracfield}
\psi_l(x,t)=\sum_{n=0}^\infty \frac 1{\sqrt{L}}\left[e^{-\frac{i2\pi}L(x^\pm-i\epsilon)(n+\frac 12)}b^\pm_n+ e^{\frac{i2\pi}L(x^\pm+i\epsilon)(n+\frac 12)}{d^\pm_n}^\dagger\right],
\ee
where $x^+$($x^-)$ goes with $\psi_1$($\psi_2$), according to the equations of motion $\partial_\mp \Psi_l=0$. The canonical anti-commutation relations require that $\{b^\pm_n,{b^\pm_m}^\dagger\}=\{d^\pm_n,{d^\pm_m}^\dagger\}=\delta_{m,n}$ and
the energy values turn out to be given by $E_n=\frac{2\pi}L (n+\frac 12)$.

The two-point functions are

\be\label{diractwopoint}
S(\Delta x^\pm)\equiv <0|\psi_l(x)\psi_l^\dagger(0)|0>=<0|\psi_l^\dagger(x)\psi_l(0)|0>=\frac 1L\sum_{n=0}^\infty e^{-\frac{2\pi (x^\pm-i\epsilon)(n+\frac 12)}L}=\frac 1{2Li\;\sin (\frac \pi L(x^\pm-i\epsilon))},
\ee
and the anti-commutation relations read

\be\label{ar}
\left\{  \psi_l(x^\pm),\psi_{l^\prime}^\dagger(x^{\pm\prime})\right\}=\delta_{l,l^\prime}\delta^A_L(x^\pm-x^{\pm\prime}),
\ee
where $\delta^A_L$ represents the  anti-periodic Dirac delta function, $ \delta^A_L(x+L)=-\delta^A_L(x)$.

A nice expression in terms of integral representation can be obtained, as shown in the appendix A,  

\bear\label{diracintegral}
<0|\psi_l(x)\psi^\dagger_l(0)|0>\!-\!<\psi_l({ { L}\over 2})\psi_l^\dagger(0)>&=&\frac 1{2\pi}\int_0^\infty dp\left[e^{-ip(x^\pm-i\epsilon)}\frac 1{1+e^{-iLp}}+e^{ip(x^\pm+i\epsilon)}\frac {e^{-iLp}}{1+e^{-iLp}}- \right.\nonumber\\
&&-\left.e^{-ip\frac {L}2}\frac 1{1+e^{-iLp}}-e^{ip\frac L2}\frac {e^{-iLp}}{1+e^{-iLp}}\right]
+\frac 12C(L,x)
.\ear
The  integral representations can be obtained through field-theoretic methods as we will explore elsewhere.

\section{ Spatially Periodic Fields Bosonization}

\setcounter{equation}{0}
Let us start with the zero temperature fields, eq. (\ref{me}) and eq. (\ref{diracfield}).

The Mandelstam operators allow to describe the fermionic functions in terms of
the bosonic ones in the case of infinite space fields. We want them to be generalized to the periodic case.
 Let us try to construct the operator as follows
 
\be\label{M15}
{\Psi_M}_\alpha(x^\pm)= \sqrt{\frac 2L} {\cal P}(x^\pm)  :e^{i2\sqrt{\pi} {\gamma_5}\phi(x^\pm;L)}:
\ee
where the Dirac index, $\alpha$, implicitly labels the diagonal component of $\gamma_5=\gamma_0\gamma_1$ and ${\cal P}(x^\pm) $ represents a pre-factor so that

$$
<\Psi_M(x^\pm)\Psi^\dagger _M(x^{\prime\pm})>=\frac 2L {S^0_M}(\Delta x^\pm)  e^{4{\pi} <0|\phi(x^\pm;L)\phi(x^{\prime\pm};L)|0>}\equiv S_M(\Delta x^\pm),
$$
where ${S^0_M}(\Delta x^\pm)$ collects the contribution from the pre-factor.   Using the eq. (\ref{discrete1}) we obtain that

  $$
  S_M(x^\pm)=\frac 2L  {\cal F}(\Delta x^\pm) e^{-\ln\left(1-e^{-2\epsilon-\frac{i2\pi \Delta x^\pm}{L}}\right)}=\frac {{\cal F}(\Delta x^\pm)e^{\epsilon+i\frac\pi L\Delta x^\pm}}{2iL\sin{(\frac \pi L(x^\pm-i\epsilon))}}.
  $$
	
This can be compared with the result in eq. (\ref{diractwopoint}). In order for the bosonic description to agree with the fermionic one we shall require ${S^0_M}(\Delta x^\pm)=e^{-\frac{i\pi \Delta x^\pm}{L}}$. The tempting definition of a c-number pre-factor, ${\cal P}(x^\pm)=? e^{-\frac{i\pi  x^\pm}L}$, would not work for the reversed order two-point function, $ <\Psi^\dagger(x^\pm)\Psi(x^{\prime\pm})>$. In order to include the description of this two-point function in the formalism  an operator pre-factor is needed. An economic procedure, which stresses the similarities with the infinite space case, is obtained from the exponentiation of a  new field $\phi^0$ with the property that

\be\label{phizero}
<0|\phi^0(x^\pm)\phi^0(x^{\prime\pm})|0>=\frac{-i}{4L}(x^\pm-x^{\prime\pm})-\frac {\ln{ \mu L}}{4\pi}.
\ee
The Hermitian field, $\phi^0$,  can be defined through

\be\label{phizeromodes}
\phi^0(x^\pm)=\frac 1{2\sqrt{2}}(c_2^\pm-c_1^\pm+c_2^{\pm\dagger}-c_1^{\pm\dagger})+\frac{ix^\pm}{2\sqrt{2}L}(c_1^\pm+c_2^\pm-c_1^{\pm\dagger}-c_2^{\pm\dagger})+\sqrt{\frac {-\ln{ \mu L}}{4\pi}}(c_3^\pm+c_3^{\pm\dagger}) 
\ee
where, for left and right modes independently, $[c_i^\pm,c_j^{\pm\dagger}]=(-1)^{(i+1)}\delta_{i,j}$, so that $c_{2}^\dagger$ creates negative norm states while $c_1^\dagger$ and $c_3^\dagger$ create positive norm ones. Anyway, the vacuum will obey $c_i|0>=0$. The term $\frac {\ln \mu L}{4\pi}$ in eq. (\ref{phizero}), and thus the $c_3$ modes in eq. (\ref{phizeromodes}), have been introduced to enforce the charge and chirality selection rules, which do not appear naturally in the periodic case, in contrast to the infinite space field case. The infinitesimal parameter $\mu$ will be taken to zero eventually.

The complete scalar field constructed from the addition of the normal mode and zero mode fields

\be\label{Phi}
\Phi(x)=\phi(x)+\phi^0(x)
\ee
 strictly obeys the canonical commutation relation 
\be\label{crP}
\left[\Phi(x^\pm),\dot{\Phi}(x^{\prime\pm})\right]=\frac i2 \delta_L(\Delta x^\pm)
\ee
and has the two point function 

\be\label{tpP}
<0|\Phi(x^\pm)\Phi(x^{\prime\pm})|0>=\frac{-1}{4\pi}\ln{\left(i\sin \left(\frac{\pi} L(\Delta x^\pm -i\epsilon)\right)\right)},
\ee
with $|0>$ now representing the vacuum of both zero and normal modes. 
The $\Phi$ field fails, however, due to the $\phi^0$ contribution, to satisfy strictly as an operator identity the periodicity in the space variable\footnote{ A more economic approach is provided by identifying both $c_1^+$ and $c_2^+$ with $c_1^-$ and $c_2^-$, so that $<0|\phi^0(x^+)\phi^0(x^{\prime -})|0>=\frac{-i}{4L}(x^+-x^{\prime -})$. The resulting field $\phi^0(x^+)+\phi^0(x^-)$ turns out not to depend on $x^1$ and, thus, $\phi^0(x)+\phi(x)$ results to be periodic. Since the $c_3$ modes are independent, the chiral selection rule is, nevertheless, enforced to the Mandelstam operator.}. As the eq. (\ref{tpP}) shows,
since the argument of the logarithm encircles the origin when $x^1\rightarrow x^1+2L$, the two-point function of $\Phi$ changes by $1\over 2$ when $x^1\rightarrow x^1+2L$. Let us call attention to the zero mode character of the $\phi^0$ field. Indeed the $\Phi$ mode expansion could be derived by adding two independent contributions to the scalar field mode expansion of the $\phi$ field, eq. (\ref{me}), both corresponding to $k\rightarrow 0$, and quantized with opposite norms. The $\phi^0$ field two-point contribution would be retrieved in the limiting case. The necessity of introducing zero modes has been noticed by Haldane \cite{haldane} a long time ago. Our treatment differs, however, in providing a framework in which the zero mode fields are described in terms of creation-annihilation operators. In order to obtain such description, the Hilbert space in which the creation-annihilation operators act is endowed with indefinite metrics. With this approach, a closer parallelism between the treatments of the infinite and the finite-periodic space is achieved. 

Note that the two-point functions of the chiral scalar fields and the ones of the Dirac fields in periodic space are related through analytic continuations to the corresponding two-point functions of the fields at infinite space and finite temperature \cite{ABR}, $\sin \left(\frac{\pi} L(\Delta x^\pm -i\epsilon)\right)\rightarrow\sinh \left(\frac{\pi} \beta(\Delta x^\pm -i\epsilon)\right)$.

The Mandelstam operators, which allow to reproduce the N-point functions of the fermion fields, result to be given by

\be\label{M15new}
\Psi_M(x^\pm)= \sqrt{ {L\mu}}   :e^{i2\sqrt{\pi}\gamma_5\left(\phi^0(x^\pm)+ \phi(x^\pm)\right)}:=\sqrt{ {L\mu}}   :e^{i2\sqrt{\pi}\gamma_5( \Phi(x^\pm))}:.
\ee

Note that in contrast to the non-periodic case, in the periodic case an infra-red regulator parameter does not emerge naturally which leads to the charge super-selection rule, $<\Psi_M(x^\pm) \Psi_M (x^{\prime\pm})>=0$. Nevertheless, a natural infra-red regulator emerges in the limit of $L\rightarrow \infty$, since
$$
<\phi(x^\pm)\phi(x^{\prime\pm})>_{L\rightarrow\infty} \approx \frac{-1}{4\pi}\ln {(2\epsilon-i2\pi x^\pm /L)},
$$
so that $1/L$ plays the role, in this limit, of an infra-red parameter, which enforces the charge selection rule, without the need
of introducing the $d_3$ modes.

In order to obtain the anti-commutation relations between distinct Dirac field components the fields shall be redefined
with Klein factors. An implementation of the Klein factor, which is smooth for $-\frac L2<x^1< \frac L2$, is given with the  Mandelstam field redefined by ( ${\gamma_5}_{1,1} =1$)

\be\label{M16}
{\Psi_{M\!K}} (x)= \sqrt{\frac {2\mu}L} :e^{i\sqrt{\pi} (\gamma_{5}\Phi(x)-\int_{x^1}^{L(1+\mbox{\tiny
 \it Int}({x^1}/L+ 1/2))} \dot\Phi (\chi)d\chi^1)}:,
\ee
where $\mbox{\it Int}(x)$ represents the integer part of $x$, indeed the greater relative number smaller or equal to  $x$. With this definition, the anti-periodicity for the fermionic field is obtained in a weak sense: the n-point functions are anti-periodic.

The prescription for the bosonization of the currents turns out to be given by

\be\label{fcurr}
J^{F}(x^\pm)=:\Psi^\dagger(x^\pm)\Psi(x^\pm):\longleftrightarrow J^{B}(x^\pm)=\frac{e^{-i\pi \over 4}}{\sqrt{2\pi}}\partial_\pm\phi(x^\pm),\ee
where a zero norm contribution from the zero-mode fields has been discarded.
The computation of the second derivative of the expression in eq. (\ref{discrete1}) establishes that the N-point function of the bosonized currents agree with the ones of the fermionic currents under this prescription.

\subsection{Symmetry aspects}

We want here to access the symmetry aspects of the bosonized fermion field. Consider first the bosonic fields.
Since the $\Phi$ field is massless it is subject to symmetry under rigid translation of the field. The   operators,
with normalization defined for future convenience,

\bear\label{symboson} 
{\Pi_0}^\pm&=&\frac 1{2\sqrt \pi}\int_{-{L\over 2}}^{L \over 2}\partial_\pm\Phi dx^1=\frac 1{2\sqrt \pi}\int_{-{L\over 2}}^{L \over 2}\left(\partial_\pm\phi+\partial_\pm\phi^0\right) dx^1=\frac 1{2\sqrt \pi}\int_{-{L\over 2}}^{L \over 2}\partial_\pm\phi^0  dx^1\nonumber\\
&=&{\frac {i}{2\sqrt{2\pi}}}\left(c_1^++c_2^+-c_1^{+\dagger}-c_2^{+\dagger}\right),
\ear
are the generators of rigid translations 
\be
\left[{\Pi_0}^\pm,\phi^0(x^\pm)\right]=\left[{\Pi_0}^\pm,\Phi(x^\pm)\right]=-\frac i {2\sqrt \pi},
\ee
while $\phi$ commutes with the generators. The zero mode field $\phi^0$ endows the quantum scalar field with this symmetry transformation.

From here we can define the charge, $Q$, and the chirality, $Q_5$, operators acting on the bosonized fermionic fields, and which could also be read from the
eq. (\ref{fcurr}),

\be
Q={\Pi_0}^+-{\Pi_0}^-\hskip 1cm\mbox{and}\hskip 1cm Q_5={\Pi_0}^++{\Pi_0}^-.
\ee
It results that
\be
e^{i\alpha Q}\Psi_M e^{-i\alpha Q}=e^{i\alpha} \Psi_M \hskip 1cm\mbox{and}\hskip 1cm e^{i\alpha Q_5}\Psi_M e^{-i\alpha Q}=e^{i\alpha\gamma_5} \Psi_M.
\ee

One sees that the zero mode field, $\phi^0$, plays the structural role of allowing for the definition of the charge and chirality of the bosonized fermion. The infrared $\mu$-factor is responsible for the implementations of the corresponding selection rules.

 \section{ Finite Temperature Fields on Compact Spaces}
\setcounter{equation}{0}
\subsection{Scalar field}
We consider here the computation of the scalar field two-point functions in the presence of a heat bath. We start by
the extension of eq. (\ref{discrete1}) in order to take account of the heat bath with inverse temperature $\beta$. As a guide line to the construction we will require that the field satisfy the KMS conditions \cite{KMS}. The two-point function will be defined by:

\be \label{discrete2}
D(\Delta x^\pm;\beta)\equiv<\phi(x^\pm)\phi(x^{\prime\pm})>_\beta=\sum_{k=1}^\infty \frac 1{4\pi k}\left[e^{-\frac{i2\pi k \Delta x^\pm }{L}}\times \frac 1{1-e^{-\frac{2\pi\beta k}L}}
 +e^{+\frac{i2\pi k \Delta x^\pm}{L}}\times \frac {e^{-\frac{2\pi\beta k}L}}{1-e^{-\frac{2\pi\beta k}L}}\right].
\ee

The sum above can be expressed in term of the theta functions, \cite{MO}, page. 375. Indeed, the regularization of the sum by making $\Delta x^\pm\rightarrow\Delta x^\pm-i\epsilon$ leads to
\bear \label{exact1}
D(\Delta x^\pm-i\epsilon;\beta)&=&\sum_{k=1}^\infty \frac 1{4\pi k}\left[e^{-\frac{i2\pi k (\Delta x^\pm-i\epsilon ) }{L}}
 +2\cos{\frac{2\pi k (\Delta x^\pm-i\epsilon)}{L}}\times \frac {e^{-\frac{2\pi\beta k}L}}{1-e^{-\frac{2\pi\beta k}L}}\right].   \nonumber\\
&=&-\frac 1{4\pi}\left[\ln\left(\gamma^{-4}ie^{-i\frac{\pi}L(\Delta x^\pm-i\epsilon)}{{\theta_1}}(\frac{\pi}L(\Delta x^\pm-i\epsilon),e^{-\frac{\pi\beta}L})\right)\right],
\ear
where $\gamma$ is de Euler-Mascheroni number.

This expression allows one to discuss the periodicity in space and time of the two-point function. The (anti)periodicity of the theta-function, $\theta_1(u+\pi,q)=-\theta_1(u+\pi,q)$, \cite{gradsh}, leads to the result that the two-point function is periodic in space, $x^1\rightarrow x^1+L$, with period L. On the other hand, the quasi-periodicity property,
$\theta_1(u-i\pi\beta/L,e^{-\frac{\pi\beta}L})=-e^\frac{\pi\beta+2iu}L\theta_1(u,e^{-\frac{\pi\beta}L})$, \cite{gradsh}, leads to the behaviour under imaginary time translations,  $x^0\rightarrow x^0-i\beta$, so that

\be\label{KMSb1}
D(\Delta x^\pm-i\beta;\beta)=D(-\Delta x^\pm;\beta).
\ee
This is the KMS condition we deserved the two-point function to satisfy. 

Furthermore, the temperature dependent two point
function can be obtained within a thermofielddynamics formalism, \cite{TFD,ABR}. First, the fields creation and annihilation operators are doubled with the introduction of new independent fields, $\{a_k\}\rightarrow\{a_k,\widetilde {a}_k\}$ and $\{\phi\}\rightarrow\{\phi,\widetilde\phi\}$. The $\widetilde\phi$ field mode expansion reads

\be \label{metilde}
\widetilde\phi(x^\pm)=\sum_{k=1}^\infty \frac{1}{2\sqrt{\pi k}}\left[ e^{i\frac{2\pi k}{L}x^\pm}e^{-\epsilon k} \widetilde a_k^{\pm}+e^{-i\frac{2\pi k}{L}x^\pm} e^{-\epsilon k} \widetilde a_k^{\pm\dagger}\right];
\ee 
A new Hilbert space is constructed as the direct product of the Hilbert spaces associated to the original mode operators and to the new tilded-operators;  The temperature dependent operators, which generalize the expression in (\ref{me}), are introduced as

\be \label{mebeta}
\phi(x^\pm;\beta)=\sum_{k=1}^\infty \frac{e^{-\epsilon k}}{2\sqrt{\pi k}}\left[ e^{-i\frac{2\pi k}{L}x^\pm} \left[ \cosh(\theta_k)a_k^{\pm}-\sinh(\theta_k){\widetilde a}_k^{\pm\dagger}\right]+e^{i\frac{2\pi k}{L}x^\pm}  \left[ \cosh(\theta_k)a_k^{\pm\dagger}-\sinh(\theta_k){\widetilde a}_k^{\pm}\right]\right],
\ee 
where $\sinh^2(\theta_k)=\frac {e^{-\frac{2\pi\beta k}L}}{1-e^{-\frac{2\pi\beta k}L}}$. 
In this language we have 

\be
D(\Delta x^\pm;\beta) = <0,0|\phi(x^\pm;\beta)\phi(x^{\prime\pm};\beta)|0,0>,
\ee
so that the KMS condition is rewritten as

\be\label{KMSb2}
<0,0|\phi(x^\pm-i\beta;\beta)\phi(x^{\prime\pm};\beta)|0,0>=<0,0|\phi(x^{\prime\pm};\beta)\phi(x^{\pm};\beta)|0,0>,
\ee
where $|0,0>$ is the vacuum state of the resulting Hilbert space. In this vacuum we have $a_k^\pm|0,0>={\widetilde a}_k^\pm|0,0>=0$. The possibility of expressing the finite-temperature two point function in terms of a vacuum expectation value has far reaching consequences as we will see later.

For completeness we present the tilded field operator mode expansion, obtained with the tilde conjugation rule,
$a\rightarrow \widetilde a$, and $c\rightarrow c^\ast$,

\be \label{mebetatilde}
\widetilde\phi(x^\pm;\beta)=\sum_{k=1}^\infty \frac{e^{-\epsilon k}}{2\sqrt{\pi k}}\left[ e^{i\frac{2\pi k}{L}x^\pm} \left[ \cosh(\theta_k){\widetilde a}_k^{\pm}-\sinh(\theta_k){ a}_k^{\pm\dagger}\right]+e^{-i\frac{2\pi k}{L}x^\pm}  \left[ \cosh(\theta_k){\widetilde a}_k^{\pm\dagger}-\sinh(\theta_k){a}_k^{\pm}\right]\right].
\ee 

From here, along the same lines that lead to the equation \ref{exact1}, it results the two-point function

\be\label{exact2}
<0,0|\widetilde\phi(x^\pm;\beta)\widetilde\phi(x^{\prime\pm};\beta)|0,0>= \left(D(\Delta x^\pm;\beta)\right)^\ast, 
\ee
and the crossed function
\be\label{exact3}
<0,0|\widetilde\phi(x^\pm;\beta)\phi(x^{\prime\pm};\beta)|0,0>=
\frac 1{4\pi}\left[\ln{\left(\gamma^{-1}\theta_4(\frac\pi L\Delta x^\pm,e^{-\frac{\pi\beta}L})\right)}\right]
\equiv \widetilde D(\Delta x^\pm;\beta).
\ee
The overall signal in this equation when compared to the analogous in eqs. (\ref{exact1}) and (\ref{exact2}) will
be important for the bosonization.

It remains to consider the temperature dependence of the contribution of zero modes. We will discuss it shortly.

 \subsection{ Finite Temperature Dirac Field on Compact Spaces}

 Our start point will be to extend (\ref{diractwopoint}) to finite temperature as
 
 \bear \label{diract0}
 <\psi_\alpha(x)\psi_\alpha^\dagger (0)>_\beta&=&<\psi_\alpha^\dagger(x)\psi_\alpha (0)>_\beta\nonumber\\
&=&\sum_{n=0}^\infty \frac 1L\left[e^{-i\frac{2\pi}L(x-i\epsilon)(n+\frac 12)}\frac{1}{1+e^{-\frac{\pi\beta}L(2n+ 1)}}+e^{i\frac{2\pi}L(x+i\epsilon)(n+\frac 12)}\frac{e^{-\frac{\pi\beta}L(2n+1)}}{1+e^{-\frac{\pi\beta}L(2n+1)}}\right],
 \ear
 where  $x=x^+$($x=x^-$) for $\alpha=1$($\alpha=2$).

Expanding the statistical factor in series, performing the sum in {\it n} first, and using \cite{gradsh}, pg. 913, we obtain the result expressed in terms of the Jacobi elliptic functions

\be\label{exactf1}
S(\Delta x^\pm;\beta)=<\psi(x^\pm)\psi^\dagger (0)>_\beta=-\frac{k{\bf K}}{\pi L} \mbox{cn}(2{\bf K}\frac {x^\pm}{L}-i{\bf K^\prime}),
\ee
where $\beta=\frac{L{\bf K^\prime}}{\bf K}$. Using \cite{gradsh}, pgs. 923 and 922, it can be re-expressed as

\be\label{diract}
S(\Delta x^\pm;\beta)=-\frac{i\sqrt{k^\prime k}{\bf K}}{\pi L} \frac{{\theta_3}(\frac {\pi x^\pm}{L},e^{-\frac{\pi\beta}{L}})}{{\theta_1}(\frac {\pi x^\pm}{L},e^{-\frac{\pi\beta}{L}})}.
\ee
Note that the numerical factor can be rewritten with  $\frac{2\sqrt{k^\prime k}{\bf K}}{\pi }=\theta_3(0)\theta_3(-\frac\pi 2)$.

In this later form, due to the (quasi)periodicity of the Jacobi
theta functions, \cite{gradsh} pgs 921 and 922,
 the anti-symmetry, the anti-periodicity, and the KMS condition
 are obtained and expressed respectively as

\bear
S(-\Delta x^\pm;\beta)&=&-S(\Delta x^\pm;\beta);\nonumber\\
S(\Delta x^\pm\pm L;\beta)&=&-S(\Delta x^\pm;\beta)\hskip 1cm \mbox{and}\nonumber\\
S(\Delta x^\pm-i\beta;\beta)&=&S(-\Delta x^\pm;\beta).
\ear

In order to discuss the limit of the two-point function when the parameter L goes to infinity it is useful
 to reconsider the expansion of the statistical factors in powers of $e^{i\frac{2\pi}L(x+i\epsilon)(n+\frac 12)}$  in the equation 
\ref{diract0} so that

\bear\label{diracdiscrete}
 S(\Delta x^\pm;\beta)&=&\sum_{n=0}^\infty \frac 1L\left[e^{-i\frac{2\pi}L(x^\pm-i\epsilon)(n+\frac 12)}+\sum_{l=1}^\infty(-1)^{l}\left[e^{-i\frac{2\pi}L(x^\pm-i\beta l)(n+\frac 12)}-e^{i\frac{2\pi}L(x\pm+i\beta l)(n+\frac 12)}\right]\right]\nonumber\\
 &=&\frac 1{2iL}\left[\frac 1{\sin{\frac \pi L(x^\pm-i\epsilon)}}+\sum_{l=1}^\infty (-1)^l\left(\frac 1{\sin{\frac \pi L(x^\pm-il\beta)}}+\frac 1{\sin{\frac \pi L(x^\pm+il\beta)}}\right)\right].
 \ear
 The sum of the limit when L goes to infinity of the last expression leads to

\bear
 S(x^\pm;\beta)_{L=\infty}&=&\frac 1{2i\pi}\left[\frac 1{{(x^\pm-i\epsilon)}}+\sum_{l=1}^\infty (-1)^l\left(\frac 1{{(x^\pm-il\beta)}}+\frac 1{{ (x^\pm+il\beta)}}\right)\right]\nonumber\\
 &=&\frac 1{2 i\beta \sinh{(\frac{\pi x^\pm}\beta)}},
 \ear
which is in accordance to the results obtained for the infinite space finite temperature Dirac field in \cite{ABR2}. 
 Besides, the eq. (\ref{diracdiscrete}) allows one to see that the singularity associated to $x\rightarrow 0$ is dominated by the first term in the sum, leading to the proper inhomogeneous term in the anti-commutation relations for the fermionic field.

The N-point fermionic functions can be obtained within Thermofielddynamics procedure. The field annihilation and creation operators are doubled $\{b_k,d_k\}\rightarrow\{b_k,d_k,\widetilde b_k,\widetilde d_k\}$, with the added anticoutation relations $\{\widetilde b_n,\widetilde b_m^\dagger\}=\{\widetilde d_n,\widetilde d_m^\dagger\}=\delta_{m,n}$. The Hilbert space is defined as the product of the Hilbert spaces where tilded and non-tilded fields act. The temperature
dependent field is defined as

\bear\label{diracfieldT}
\Psi_l(x,t;\beta)&=&\sum_{n=0}^\infty \frac 1{\sqrt{L}}\left[e^{-\frac{i2\pi}L(x^\pm-i\epsilon)(n+\frac 12)}\left(\cos{\theta^F_n}b^\pm_n+i\sin{\theta^F_n}{\widetilde b}^\pm_n{}^\dagger\right)+ \hskip 2cm\right.\nonumber\\
&&\hskip 2cm\left.e^{\frac{i2\pi}L(x^\pm+i\epsilon)(n+\frac 12)}\left(\cos{\theta^F_n}{d^\pm_n}^\dagger-i\sin{\theta^F_n}{\widetilde d}^\pm_n\right)\right],
\ear
where $\sin^2{\theta^F_n}=\frac{e^{-\frac{2\pi\beta}L(n+1/2)}}{1+e^{-\frac{2\pi\beta}L(n+1/2)}}$. 
In this language

$$
S(\Delta x^\pm;\beta)=<0,0|\Psi_\alpha(x;\beta)\Psi_\alpha^\dagger (0;\beta)|0,0>.
$$
The fermion n-point functions are obtained from the (Fermion)vacuum expectation values of thermal fields, analogously to the normal-mode scalar field.

The tilded fermion field is given by

\bear\label{diracfieldTtilded}
\widetilde\Psi_l(x,t;\beta)&=&\sum_{n=0}^\infty \frac 1{\sqrt{L}}\left[e^{\frac{i2\pi}L(x^\pm+i\epsilon)(n+\frac 12)}\left(\cos{\theta^F_n}\widetilde b^\pm_n-i\sin{\theta^F_n}{ b^\pm_n}^\dagger\right)+\hskip 2cm\right.\nonumber\\
&&\hskip 2cm\left. e^{\frac{-i2\pi}L(x^\pm-i\epsilon)(n+\frac 12)}\left(\cos{\theta^F_n}{\widetilde d^\pm_n}{}^\dagger+i\sin{\theta^F_n}{ d}^\pm_n\right)\right],
\ear

From here, along the same lines described above, the tilded two-point functions are derived

\be\label{diract2}
<0,0|\widetilde\Psi_\alpha(x;\beta)\widetilde\Psi_\alpha^\dagger (0;\beta)|0,0>=\frac{k{\bf K}}{\pi L} \mbox{cn}(2{\bf K}\frac {x^\pm}{L}-i{\bf K^\prime})=(<0,0|\Psi_\alpha(x;\beta)\Psi_\alpha^\dagger (0;\beta)|0,0>)^\ast,
\ee
while the crossed functions are

\be\label{diractcrossed}
<0,0|i\widetilde\Psi_\alpha(x;\beta)\Psi_\alpha (0;\beta)|0,0>=-\frac{k{\bf K}}{\pi L} \mbox{cn}(2{\bf K}\frac {x^\pm}{L})=-\frac{\sqrt{k^\prime k}{\bf K}}{\pi L} \frac{{\theta_2}(\frac {\pi x^\pm}{L},e^{-\frac{\pi\beta}{L}})}{{\theta_4}(\frac {\pi x^\pm}{L},e^{-\frac{\pi\beta}{L}})}.
\ee
Note that the tilded field charge charge is opposed to the non-tilded field one. From the first of the above equations we see that

\be
<0,0|i\widetilde\Psi_\alpha(x;\beta)\Psi_\alpha (0;\beta)|0,0>=-<0,0|\Psi^\dagger_\alpha(x-i\frac\beta 2;\beta)\Psi_\alpha (0;\beta)|0,0>.
\ee
This relationship is also witnessed in the infinite space case \cite{ABR2}.

In the sequel we will derive not only the eq. (\ref{diract}) but also the eqs. (\ref{diract2} and \ref{diractcrossed})
within the thermofieldbosonization.


\section{Non Zero Temperature Case Bosonization}
\setcounter{equation}{0}

In the following we consider bosonization in the case with non-zero temperature. We keep the eq. (\ref{M15new}) as an Ansatz,
and consider the heat bath contributions for the $\phi$ field N-point functions. As for the $\phi^0 $ field, we consider, as a working hypothesis, their N-point functions to be factorized from the ones of the $\phi$ field even in
the presence of the heat bath. It results in

\bear\label{propagatoridentification}
S_M(\Delta x^\pm;\beta)&=&<\Psi_M(x^\pm)\Psi^\dagger _M(x^{\prime\pm})>_\beta\nonumber\\
&=&{ {L\mu}}<:e^{i2\sqrt{\pi}\gamma_5(\phi^0(x^\pm)+ \phi(x^\pm))}:
:e^{-i2\sqrt{\pi}\gamma_5(\phi^0(x^{\prime\pm})+ \phi(x^\pm))}:>_\beta\nonumber\\
&\equiv&S^0_M(\Delta x^\pm;\beta)\;<:e^{i2\sqrt{\pi}\gamma_5\phi(x^\pm;L)}::
e^{-i2\sqrt{\pi}\gamma_5\phi(x^{\prime\pm};L)}:>_\beta\nonumber,
\ear
where $S^0_M(\Delta x^\pm;\beta)$ collects the finite-temperature contribution of the zero modes.

Using the representation in eq. (\ref{mebeta}) and the equation (\ref{exact1}), we see that the contribution
of the normal modes at finite temperature leads to

\bear
S_M(\Delta x^\pm;\beta)&=&S^0_M(\Delta x^\pm;\beta)
e^{4{\pi} <0,0|\phi(x^\pm;\beta)\phi(x^{\prime\pm};\beta)|0,0>}\nonumber\\
&=&S^0_M(\Delta x^\pm;\beta)     \frac{ \gamma^{4}e^{i\frac{\pi\Delta x^\pm}L}  }{i\theta_1(\frac{\pi\Delta x^\pm}L,e^{-\frac{\pi\beta}L})}.
\ear
The comparison  of this last expression to the fermionic two-point function,  eq. (\ref{diract}), 
requiring that $S_M(\Delta x^\pm;\beta)=S(\Delta x^\pm;\beta)$, what is needed for
the bosonization
of the space periodic and finite temperature free fields to work as an exact mapping, requires the identification

\be\label{presc}
 S^0_M(\Delta x^\pm;\beta) =\frac{\sqrt{k^\prime k}{\bf K}}{\gamma^{4}\pi L}e^{-i\frac{\pi\Delta x^\pm}L}  \theta_3(\frac {\pi x^\pm}{L},e^{-\frac{\pi\beta}{L}}) .  
\ee
This means a prescription for the finite temperature computation of the zero mode contributions to the
bosonic two-point function. This prescription can be understood in two alternative forms.

The first interpretation results from the series expansion, \cite{gradsh},
 $\theta_3(\frac {\pi x^\pm}{L},e^{-\frac{\pi\beta}{L}})=\sum_{n=-\infty}^{\infty} e^{-\frac{\pi\beta n^2+2i \pi nx^\pm}L}$, which leads to

$$
S^0_M(\Delta x^\pm;\beta)={\cal N}(\beta)e^{-i\frac{\pi\Delta x^\pm}L} \sum_{n=-\infty}^\infty e^{-\frac\pi L(\beta n^2-2ix^\pm n)}.
$$
This expression, as is shown in the appendix, is obtained from the zero mode field as the result of the computation

\be\label{zmtopological}
S^0_M(\Delta x^\pm;\beta)={\cal N}(\beta)e^{-i\frac{\pi\Delta x^\pm}L} \sum_{n=-\infty}^\infty <0|{U_0^{\pm}}^{\dagger n}: e^{i2\sqrt{\pi}(\phi^0(x^\pm)-\phi^0(x^{\prime\pm}))}:e^{-\beta H_0}
{U_0^{\pm}}^n|0>,
\ee
where $H_0$ is the Hamiltonian for the zero mode field
$$
H_0=\frac{\pi}{L}\left({{\Pi_0}^+}^2+{{\Pi_0}^-}^2\right),
$$
and where ${\Pi_0}^\pm$ have been introduced in (\ref{symboson}),
and in which 

\be\label{ladder}
U_0^{\pm}\equiv e^{i2\sqrt{\pi}\phi^0(x^\pm=0;L)},
\ee computed at $x^\pm=0$, acts as ladder operator for ${\Pi_0}$. The interpretation is ultimately in terms of the sum of energy values.

This observation prompts to the
construction of the zero mode fields in the thermofielddynamics approach. It is economic to define, for each chiral component independently,  $q=\sqrt{\frac \pi2}(c_2-c_1+c_2^\dagger-c_1^\dagger)$, so that, here $x=x^\pm$ and the corresponding chiral index of the creation and annihilation operators is omitted, 

$$\phi^0(x)=\frac 1{2\sqrt \pi}q+\sqrt\pi\frac {x}{ L}\Pi_0+ \frac f{\sqrt 2}(c_3+c_3^\dagger).$$
Besides the doubling the normal modes fields,
as we have already discussed, the zero mode fields are also doubled, $\{\phi^0\}\rightarrow \{\phi^0,\widetilde\phi^0\}$, with $\{q,\Pi_0,c\}\rightarrow\{q,\Pi_0,c,\widetilde q,{\widetilde \Pi}_0,\widetilde c\}$, such that

$$\widetilde\phi^0(x)=\frac 1{2\sqrt \pi}\widetilde q+\sqrt\pi\frac {x}{ L}\widetilde\Pi_0+ \frac f{\sqrt 2}(\widetilde c_3+\widetilde c_3^\dagger).$$
Observe the new commutation relation $[\widetilde q,\widetilde \Pi_0]=-i$, according to the tilde-conjugation rule..

 The states $|n>\equiv U_0^n|0>$  are introduced and generalized to product of states of non-tilded and tilded states, $|n^\pm>\rightarrow|n^\pm,m^\pm>=|n^\pm>\otimes|m^\pm>$. This construction is detailed in the appendix B. The thermal expected value of the zero modes turns out to be described as expected values of the zero mode fields on
the theta-state, the thermal-ground state ( not thermal vacuum!)

$$
|\theta^\pm>\propto \sum_{n^\pm=-\infty}^\infty e^{-\beta E_n/2}|n^\pm,n^\pm>.
$$ 
With this doubling of the Hilbert space of each independent chiral fields, the thermofielddynamics description of
the model at finite temperature is constructed.  For the zero modes we obtain

$$ 
 S^0_M(\Delta x^\pm;\beta)=L\mu<\theta^\pm|: e^{i2\sqrt{\pi}\phi^0(x^\pm)}:: e^{-i2\sqrt{\pi}\phi^0(x^{\prime\pm})}:|\theta^\pm>.
$$

In the thermofielddynamics formalism the two-point functions are promoted to a matrix structure, so that the argument
in the eq. \ref{propagatoridentification} has to be revisited not only with regards to the tilde-tilde components,
but also with regards to the non-trivial non-tilde-tilde components. As we show in the appendix B these functions
are also retrieved with the formalism here developed. We obtain that

$$ 
 \left(S^0_M(\Delta x^\pm;\beta)\right)^\ast=L\mu<\theta^\pm|: e^{i2\sqrt{\pi}\widetilde\phi^0(x^\pm)}:: e^{-i2\sqrt{\pi}\widetilde\phi^0(x^{\prime\pm})}:|\theta^\pm>,
$$
and also that
$$ 
 \widetilde S^0_M(\Delta x^\pm;\beta)\equiv L\mu<\theta^\pm|(-i): e^{i2\sqrt{\pi}\widetilde\phi^0(x^\pm)}:: e^{i2\sqrt{\pi}\phi^0(x^{\prime\pm})}:|\theta^\pm>=-\frac{\sqrt{kk^\prime}K}{\pi L}\theta_2(\pi\frac{x^\pm}L,e^{-\pi\frac\pi\beta}),
$$

The tilded-fermi-field turns out to be given by

\be\label{M15newtilde}
\widetilde\Psi_M(x^\pm)= \sqrt{ {L\mu}}   :e^{i2\sqrt{\pi}\gamma_5\left(\widetilde\phi^0(x^\pm)+ \widetilde\phi(x^\pm)\right)}:=\sqrt{ {L\mu}}   :e^{i2\sqrt{\pi}\gamma_5( \widetilde\Phi(x^\pm))}:.
\ee

This expression is the compact-space analogous of the bosonization expression found in the infinite-space case
 in \cite{ABR2}. The signal in the exponential, which disagrees with the tilde-conjugation rules, has been discussed
in the infinite case in \cite{ABR2}. Together with

\be\label{M15new2}
\Psi_M(x^\pm)= \sqrt{ {L\mu}}   :e^{i2\sqrt{\pi}\gamma_5\left(\phi^0(x^\pm)+ \phi(x^\pm)\right)}:=\sqrt{ {L\mu}}   :e^{i2\sqrt{\pi} \gamma_5\Phi(x^\pm)}:.
\ee
they describe completely the N-point fermion functions at finite temperature, agreement with the eqs. (\ref{diract}, \ref{diract2} and \ref{diractcrossed}). The difference in the signals of the normal mode fields witnessed in the computation of tided-non-tilde functions is essential for obtaining this equivalence.

This above construction should be compared to the finite-temperature treatment of the
bosonization proposed by Haldane, \cite{haldane}, who introduced the zero modes based upon operators playing the
roles of $\Pi_0$ and ($\ln$ of) $U_0$. In the indefinite norm space formalism here presented, these operators re-appear  expressed in terms of creation an annihilation operators. The temperature dependent n-point functions which are derived from the sum over all zero mode energies,  are described here as the sum over the expected values of zero norm zero mode operators in the states $|n^+>\otimes|m^->=(U_0^+)^{n^+}(U_0^-)^{m^-}|0>\otimes|0>$, which are not, rigorously, eigenstates of the Hamiltonian. 
From the structural aspect of relativistic quantum field theory the observables are obtained from the polynomial
algebra of fields and of their derivatives. Within this principle the $\phi^0$ fields are observables. Both $q^\pm$ and ${\Pi_0}^\pm$, which satisfy $[q,\Pi_0]=i$, turn out to be the building blocks for the construction of the zero mode sector observables. Further, the mapping can be defined $U_0^n|0>\rightarrow |n>$, so that the states $|n>$, turn out to be represented by
$<q|n>\propto e^{inq}$, which satisfy $<n|n^\prime>=\delta_{n,n^\prime}$. With this in mind the equivalence of the non-tilded sector of our construction, with the ones previously found in the literature, \cite{recentperiodic2,qedc}, should be established.  The description in terms of creation-annihilation operators, however, is more transparent for the comparison with the infinite space bosonization.


The other interpretation, for the prescription in eq.(\ref{presc}), which can be obtained from the use of the product representation of the
$\theta_3$ function, is most easily obtained by considering the following translation property of the 
Jacobi theta-functions:

$$
\theta_3(\frac{\pi x^\pm}{L},e^{-\frac{\pi\beta}L})=-e^{-\pi L(\frac \beta 2 +ix^\pm)}\theta_1(\frac{\pi x^\pm} L-\frac\pi {2L}(L+i\beta),e^{-\frac{\pi\beta}L}).
$$
This leads to the following expression 

$$
 S^0_M(\Delta x^\pm;\beta) =-e^{-i\frac{\pi\Delta x^\pm}L} \frac{\sqrt{k^\prime k}{\bf K}}{\pi L} e^{-4{\pi} <0,0|\phi(x^\pm-\frac{L+i\beta}2;\beta)\phi(x^{\prime\pm};\beta)|0,0>}.
$$
Surprisingly, the temperature dependent parcel of the zero modes  contribution is related to the expected values of the temperature dependent normal modes contribution with the argument of the field displaced by half the space and half the imaginary-time
periods. The zero-temperature contribution of the zero-modes corresponds to the first factor on the right-hand side.

The fermionic two-point functions (and thereby, through the fermionic Wick theorem, similarly the N-point ones) turn out to be given by
$$
 S_M(\Delta x^\pm;\beta) =-\frac{\sqrt{k^\prime k}{\bf K}}{\pi L}e^{-i\frac{\pi\Delta x^\pm}L}  e^{4{\pi}\left(<0,0|\phi(x^\pm;\beta)\phi(x^{\prime\pm};\beta)|0,0>- <0,0|\phi(x^\pm-\frac{L+i\beta}2;\beta)\phi(x^{\prime\pm};\beta)|0,0>\right)}.
$$
 
This later interpretation is important since the finite-temperature bosonization can now be alternatively understood in terms of 
a modified thermofielddynamics formalism. Since the normal modes can be described in the thermofielddynamics formalism,
as we have outlined in the eq. (\ref{mebeta}), the finite-temperature bosonization is recast in terms of
zero-temperature zero-modes and finite temperature modified  normal-modes. The two-point function of the normal modes, modified to include the subtraction of the translated field, turns out to be written as

$$
<0,0|\left(\phi(x^\pm;\beta)\phi(x^{\prime\pm};\beta)- \phi(x^\pm- L/2-i\beta/2;\beta)\right)\phi(x^{\prime\pm};\beta)|0,0>=\hskip 6cm$$
\be
\hskip 3cm=\sum_{k=1}^\infty\frac 1{4\pi k}\left(e^{-i2\pi k\frac{\Delta x^\pm}L}\left(\frac{1-(-1)^ke^{-\frac{\pi\beta k}L}}{1-e^{-\frac{2\pi k\beta}L}}\right)+e^{i2\pi k\frac{\Delta x^\pm}L}\left(\frac{e^{-\frac{2\pi k\beta}L}(1-(-1)^ke^{\frac{\pi\beta k}L})}{1-e^{-\frac{2\pi k\beta}L}}\right)\right).
\ee

The generalization to N-point functions is discussed in the appendix. The main result is that the normal modes
contributions, which are obtained from the vacuum expectation values of exponentials of free fields, appear directly in terms of the products of expected values of pairs of exponentials. However, the zero modes ones are obtained
from the expected values on the $\theta$ states, not on the vacuum, and appear naturally in terms
of Jacobi functions with arguments depending on the N-points.

 The computation of the N-point functions can be
proceeded in two alternative and equivalent ways: 1) In the completely bosonised way, the temperature dependent functions of the zero mode and normal mode fields
are computed independently. The normal modes factor can be computed as  vacuum expected values of thermalized fields.
The zero mode is computed with the zero temperature fields expected values on the theta-states. All this computation proceeds within the thermofielddynamics formalism. 2)  The N-point functions are reduced to sums of two-point functions using the fermionic representation. The two point functions are computed with in such a way that the zero mode contribution is computed at zero temperature.
The normal modes contribution is calculated with a renormalized two-point function which accounts for the subtraction
scheme, $<\phi(x)\phi(x^\prime)>\rightarrow <(\phi(x)-\phi(x-L/2-i\beta/2))\phi(x^\prime)>$, and which can also be obtained from a 
modified thermofielddynamics formalism.

\section{Conclusion}
\setcounter{equation}{0}

In this paper we have constructed, with canonical treatment, the scalar and the Dirac fields on compact space and have presented the representation of the fermionic one in terms of the bosonic field. The procedure emphasises the similarities between the infinite and space periodic cases by presenting the fields entirely in terms of creation- annihilation operators acting on the (Fock) vacuum. Some subtleties arise in the treatment which are absent in the infinite space models.

The zero mode field $\phi^0$, expressed in terms of creation and annihilation operators, emerges naturally in order to reproduce the fermionic n-point functions from the bosonic ones. This new field plays the role of translating the periodic boundary conditions satisfied by the scalar field into anti-periodic conditions satisfied by fermionic one. Its necessity could be anticipated, however, without resource to the requirements of bosonization, if the strict canonical commutation relations, and thereby micro-causality, are required. These are obeyed by the $\Phi=\phi+\phi^0$ field but not by $\phi$. Nevertheless strict periodicity is lost for the $\Phi$ field and not for $\phi$. It is curious that such a complementarity between periodicity and micro-causality is witnessed for the scalar field  but not for the Dirac field.

This new scalar field plays the distinctive role of allowing for the definition of the charge and chirality of the
bosonized fermion fields. In the gauged version of the free fermion model, the space periodic Schwinger model,
they will be ultimately the protagonists in the process of charge and chirality condensation in the vacuum trough
the definition of the $\theta$-vacua. This has been dealt with, in treatments which do not use thermofielddynamics, for instance, in \cite{qedc}. 

The main result has been the study of the finite-temperature versions of the models within a thermofielddynamics procedure. We have constructed the two-point functions in this case, such that they satisfy the KMS condition and have presented the complete bosonization procedure. We performed an extension of the Mandelstam\cite{M} construction of the fermion field operators in terms of the bosonic field operators, such that the bosonized description of the fermion field functions displays correctly the periodicity and the anti-commutations required. This treatment extends the thermofieldbosonization \cite{ABR,ABR2} to
periodic space. The full four-fold structure\cite{OjimaMats} of the two-point fermionic functions  ($<\psi\psi^\dagger>$,$<\widetilde\psi\widetilde\psi{}^\dagger>$,$<\psi\widetilde\psi>$ and $<\widetilde\psi{}^\dagger\psi^\dagger>$), are obtained from the analogous structure of the bosonic field,
which depends upon both normal and zero modes. In the limit of infinite space the zero modes contributions decay and
the infinite space thermofield-bosonization is obtained.

The finite temperature bosonization has been discussed with two alternative view-points. Either the zero mode temperature contributions are described in terms of expected values of the zero-mode fields on the theta-state or their temperature dependences is, surprisingly, embedded into the ones of the normal modes. 

We thank the careful reading of an early version of the manuscript by K. D. Rothe and the discussions thereby.

\appendix{{\centerline{\bf{Appendix A }}}}
\vspace{0.5cm}
\renewcommand{\theequation}{{A}.\arabic{equation}}\setcounter{equation}{0}
\centerline{\bf Integral representation of the propagators}

For the scalar field at zero temperature, we start from eq. (\ref{discrete1}) re-stated as
\be\label{twop1}
D(\Delta x^\pm)=\frac {-1}{4\pi}\left[\ln\left(e^{i\pi(\frac{ L-2\Delta x^\pm }{2L})}2\sin \left(\frac {\pi\Delta x^\pm}{L}\right)\right)-\ln\left(2\sin \left(\frac {\pi}{2}\right)\right)\right],
\ee
and use the product representation  $\sin(z)=z\prod_{k=1}^\infty \frac{k^2\pi^2-z^2}{k^2\pi^2}$.  For $0<\Delta x^\pm<L$ it results
\be
D(\Delta x^\pm) =\frac {-1}{4\pi}\left[i\frac\pi{2L}(L- 2\Delta x^\pm )+\ln 2+\sum_{k=0}^\infty \ln\frac{k+\Delta x^\pm/L}{k+1/2}+\sum_{k=1}^\infty \ln\frac{k-\Delta x^\pm/L}{k-1/2}\right],
\ee
Now since, for $x,y>0$, $ 
\int_0^\infty \frac{dp}p\left[e^{-ip(x-i\epsilon)}-e^{-ip(y-i\epsilon)}\right]=\ln \frac yx,$
the integral representation results
 
\bear
D(\Delta x^\pm) &=&\frac {-1}{4\pi}\left[ i\frac\pi{2L}( L-2\Delta x^\pm )+\ln 2+\sum_{k=0}^\infty 
\int_0^\infty \frac {dp}p\left[ e^{-ip(k+1/2-i\epsilon)}-e^{-ip(k+\Delta x^\pm/L-i\epsilon)}\right]\right.\nonumber\\
&&+\left.\sum_{k=1}^\infty 
\int_0^\infty \frac {dp}p\left[e^{-ip(k-1/2-i\epsilon)}-e^{-ip(k-\Delta x^\pm/L-i\epsilon)}\right]\right]\nonumber\\
&=& \frac{i\pi}{4L}( \Delta x^\pm -\frac L2)+\ln 2+
\int_0^\infty \frac {dp}{4\pi p}\left[\left( e^{-ip(\Delta x^\pm-i\epsilon)}\frac 1{1-e^{-ipL}}+e^{ip(\Delta x^\pm+i\epsilon)}\frac{e^{-ipL}}{1-e^{-ipL}}\right)-\right.\nonumber\\
&&\left.\left. e^{-ip(L/2-i\epsilon)}\frac 1{1-e^{-ipL}}+e^{ip(L/2+i\epsilon)}\frac{e^{-ipL}}{1-e^{-ipL}}\right)\right].
\ear

A nice expression in terms of integral representation can also be obtained to the Dirac field at zero temperature through

\bear\label{diracintegral0}
<0|\psi_l(x)\psi^\dagger_l(0)|0>\!\!&-&\!\!<\psi_l({ { L}\over 2})\psi_l^\dagger(0)>=\frac{(2iL)^{-1}}{\sin (\frac\pi L x^\pm)}-\frac{(2iL)^{-1}}{\sin (\frac\pi 2)}=\frac 1{2iL\pi}\left[\sum_{n=-\infty}^{\infty} {\!\!}^\prime\frac{(-1)^n}{\frac{x^\pm}L -n}-\sum_{n=-\infty}^{\infty} {\!\!}^\prime\frac{(-1)^n}{\frac 1 2 -n}\right]\nonumber\\
&=&\sum_{n=-\infty}^{\infty} \left[\frac{(-1)^n}{2iL(\frac\pi Lx^\pm -n\pi)}-\frac{(-1)^n}{2iL(\frac\pi 2 -n\pi)}\right]\nonumber\\
&=&\frac 1{2\pi}\int_{0}^\infty dp\left[\sum_{n=0}^\infty(-1)^n\left(e^{-ip(x^\pm+Ln-i\epsilon)}-e^{-ip(\frac L2 +Ln)}\right)
-\right.\nonumber\\
&&-\left.\sum_{n=1}^\infty(-1)^n\left(e^{ip(x^\pm-Ln+i\epsilon)}-e^{ip(\frac L2- Ln)}\right)\right]+\frac 12C(L,x)\nonumber\\
&=&\frac 1{2\pi}\int_0^\infty dp\left[e^{-ip(x^\pm-i\epsilon)}\frac 1{1+e^{-iLp}}+e^{ip(x^\pm+i\epsilon)}\frac {e^{-iLp}}{1+e^{-iLp}}- \right.\nonumber\\
&&-\left.e^{-ip\frac {L}2}\frac 1{1+e^{-iLp}}-e^{ip\frac L2}\frac {e^{-iLp}}{1+e^{-iLp}}\right]
+\frac 12C(L,x)
.\ear
Here, the prime in the first sums indicates symmetric summations for $n>0$ and $n<0$, what is evaded in the next step
due to the subtraction explicit in the left hand side. The function $C(L,x)=\sum_{l=1}^\infty \delta (x-lL)$  is the singular contribution associated to the exchange
$i\epsilon\rightarrow -i\epsilon$ that has been done in the $n<0$ terms. This integral representation can be obtained through field-theoretic methods as we will explore elsewhere.

The integral representations of the space and imaginary-time periodic functions are now obtained.
We will start from  eq. (\ref{diracdiscrete}). The contribution of the first terms on the right hand side, which do not depend on $\beta$, has already been computed in eq. ( \ref{diracintegral}). For the other terms the choice of
the signal in the exponential is dictated by the signal of $\pm i\beta$. The result will be

$$
<\psi_l(x)\psi^\dagger_l(0)>_\beta-<\psi_l(L/2)\psi^\dagger_l(0)>_\beta=
<\psi_l(x)\psi_l^\dagger(0)>-<\psi_l(L/2)\psi_l^\dagger(0)>+\sum_{l=1}^\infty\sum_{n=-\infty}^\infty
\frac{(-1)^n}{2\pi}\times 
$$

\be\label{diracTL}
\;\;\;\;\;\;\;\int_{0}^\infty dp\left[e^{-ip(x^\pm-il\beta-nL)}-e^{+ip(x^\pm+il\beta-nL)} 
-e^{-ip(\frac L2-il\beta-nL)}+e^{ip(\frac L2+il\beta-nL)}\right]
\ee

It is interesting in this point to perform the sums of negative and positive values of $n$ independently. Also, the contribution of the first two parcels are obtained from (\ref{diracintegral}) with the result

\bear\label{diracTL2}
<\psi_l(x)\psi^\dagger_l(0)>_\beta-<\psi_l(L/2)\psi^\dagger_l(0)>_\beta&=&\frac 1{2\pi i}\left[\frac 1{x^\pm-i\epsilon}-\frac 2{L-i\epsilon}\right]+\Delta F
\ear
where
\bear
\Delta F&=&\frac 1{2\pi}\int_0^\infty dp\left[\sin^2\theta_L+\sin^2\theta_\beta(1-\sin^2\theta_L-(\sin^2\theta_l)^*)\right]\times\nonumber\\
&&\left[-e^{-ip(x^\pm-il\beta-nL)}+e^{+ip(x^\pm+il\beta-nL)}+e^{-ip(\frac L2-il\beta-nL)}-e^{ip(\frac L2+il\beta-nL)}
\right],
\ear
with $\sin^2\theta_L=\frac{e^{-ilP-\epsilon}}{1+e^{-ilP-\epsilon}}$ and $\sin^2\theta_\beta=\frac{e^{-\beta p}}{1+e^{-\beta p}}$. It is interesting to note that in the limit of $\epsilon\rightarrow 0$ the sum of $\sin^2\theta_L$ with $(\sin^2\theta_L)^\ast$ implements the discretization of the momenta

$$
\lim_{\epsilon\rightarrow 0}\left(1-\sin^2\theta_L-(\sin^2\theta_L)^\ast\right) =\frac {2\pi}{L}\delta_{\frac{2\pi}{L}}(p-\frac \pi L).
$$
Here we have used the periodic delta function. This result is related to the construction of the fields in the context
of thermofielddynamics.

\appendix{{\centerline{\bf{Appendix B }}}}
\vspace{0.5cm}
\renewcommand{\theequation}{{B}.\arabic{equation}}\setcounter{equation}{0}
\centerline{\bf Zero mode contributions at Finite T}

The computation the zero-mode contributions to the two-point functions of  $\psi_M$, at finite temperature will be performed. We will consider one of the
chiral components, $x^+$ or $x^-$, and streamline the notation by withdrawing the chiral upper-script, $x^\pm\rightarrow x$, $c_i^\pm\rightarrow c_i$, etc. It is useful to define the operator $q=\qm+\qM$ with the creation and annihilation ( zero-norm) parts given as

$$\qM=\sqrt{\frac \pi 2}(c_2-c_1)\hskip 2cm \mbox{and}\hskip 2cm\qm=\sqrt{\frac \pi 2}(c_2^\dagger-c_1^\dagger)$$
and similarly $\Pi_0=\pim+\piM$, with
$$\piM =\frac i{2\sqrt{2\pi}}(c_2+c_1)\hskip 2cm \mbox{and}\hskip 2cm \pim=-\frac i{2\sqrt{2\pi}}(c_2^\dagger+c_1^\dagger),$$
such that $[q,\Pi_0]=i$, with the non zero commutation relations restricted to $[q^{(\pm)},\Pi^{(\mp)}_0]=\frac i2$.

The zero mode scalar field is given by $\phi^0(x)=\frac 1{2\sqrt \pi}q+\sqrt\pi\frac {x}{ L}\Pi_0+  \frac f{2\sqrt{\pi}}(c_3+c_3^\dagger)$, with $f=\sqrt{-\ln{L\mu}}$. Using

$$
:e^{i2\sqrt\pi\phi^0(x)}:=e^{\left(i\qm+i2\pi\frac xL\pim+ i f c_3^\dagger\right)}e^{\left(i\qM+i2\pi \frac xL\piM+ i f c_3\right)},
$$
it is easy to obtain that

\bear
:e^{i2\sqrt\pi\phi^0(x)}::e^{-i2\sqrt\pi\phi^0(y)}:&=&::e^{i2\sqrt\pi(\phi^0(x)-\phi^0(y))}:e^{-[i\qM+i{2\pi}\frac xL\piM+  f c_3,i\qm+i{2\pi}\frac yL\pim+  f c_3^\dagger]}\nonumber\\
&=&\frac 1{L\mu}:e^{i2\sqrt\pi(\phi^0(x)-\phi^0(y))}:e^{-i\pi(x-y)/L}=\frac 1{L\mu}e^{i{2\pi}\Pi_0\frac{x-y}L}e^{-i\pi(x-y)/L}.
\ear

Now, defining the ladder operator
$$
U_0=e^{i2\sqrt\pi \phi^0(x)}/_{x=0}=\sqrt{L\mu}\;e^{iq}:e^{if(c_3+c_3^\dagger)}:\equiv e^{iq}U_3,
$$
the following algebra holds

\be
U_0^\dagger{\Pi_0}U_0={\Pi_0}+1, \hskip 2cm \mbox{so that}\hskip 2cm
(U_0^\dagger)^nF(\Pi_0)(U_0)^n=F({\Pi_0}+n).
\ee
Also, since $\pim$ acting on the vacuum creates zero norm states, we have that 

\be
<0|F(\Pi_0)|0>=F(0). 
\ee
In the limit when $\mu\rightarrow 0$ the $U_3$ operator enforces  the charge selection rules 

\be\label{chargesel}
<0|{U_0^\dagger}^n{U_0}^{n^\prime}|0>=\delta_{n,n'}.\ee
 The operators $U_0$ play the roles of ladder operators and of charge holders.

Finally we obtain that

$$
\sum_{n=-\infty}^{\infty}<0|(U_0^\dagger)^n\sqrt{L\mu}:e^{i2\sqrt\pi\phi^0(x)}:\sqrt{L\mu}:e^{-i2\sqrt\pi\phi^0(y)}:e^{-\beta H_0}(U_0)^n
|0>=\hskip 5cm
$$
\bear\label{sum1}
\hskip1cm&=& e^{-i\pi\frac{\Delta x}L}\sum_{n=-\infty}^{\infty}<0|(U_0^\dagger)^n:e^{i2\pi\Pi_0\frac{x-y}L}:e^{-\beta H_0}(U_0)^n|0>\nonumber\\
&=&e^{-i\pi\frac{\Delta x}L}\sum_{n=-\infty}^{\infty} e^{2 i n\pi\frac{\Delta x}L-n^2\pi\frac\beta L}    =e^{-i\pi\frac{\Delta x}L}\theta_3(\frac{\pi\Delta x}L,e^{-\pi\frac\beta L}).
\ear
This establishes the eq. (\ref{zmtopological}), with ${\cal N}(\beta)=\frac{\theta_3(0)\theta_3(\frac\pi 2)}{2\gamma^4L}$, after restoring the chiral variables $x\rightarrow x^\pm$.

The charge selection rule eq. (\ref{chargesel}) allow to re-derive this result in the thermofielddynamics formalism.
The zero mode field is doubled with the introduction of new operators

$$\widetilde\phi^0(x)=\frac 1{2\sqrt \pi}\widetilde q+\sqrt\pi\frac {x}{ L}\widetilde\Pi_0+ \frac f{2\sqrt \pi}(\widetilde c_3+\widetilde c_3^\dagger),$$
where  $\widetilde q=\sqrt{\frac \pi 2}(\widetilde c_2-\widetilde c_1+\widetilde c_2^\dagger-\widetilde c_1^\dagger)$, and  $\widetilde\Pi_0 =\frac {-i}{2\sqrt{2\pi}}(\widetilde c_2+\widetilde c_1-\widetilde c_2^\dagger-\widetilde c_1^\dagger).$                    

In the product space of the Hilbert spaces where the tilded and non-tilded zero mode fields act the theta-state
is introduced
$$
|\theta^\pm>=N_\theta \sum_{n^\pm=-\infty}^\infty e^{-\beta E_n/2}|n^\pm,n^\pm>,
$$ 
with
$$
|n^\pm,m^\pm>={U_0^\pm}^n{\widetilde U_0^\pm}{}^m|0,0>,
$$
and where $|0,0>=|0>\otimes|0>$ is the normal modes "vacuum state". The tilded ladder operators are ( observe the failure of the tilde-conjugation rule)

$$
\widetilde U_0=e^{i2\sqrt{\pi}\widetilde \phi^0(x=0)}.$$

With this setting the content in eq. (\ref{sum1}) is re-obtained

$$
<\theta|\sqrt{L\mu}:e^{i2\sqrt\pi\phi(x)}:\sqrt{L\mu}:e^{-i2\sqrt\pi\phi(y)}:|\theta>=N_\theta^2 e^{-i\pi\frac{\Delta x}L}\theta_3(\frac{\pi\Delta x}L,e^{-\pi\frac\beta L}),
$$
as well as
$$
<\theta|\sqrt{L\mu}:e^{i2\sqrt\pi\widetilde\phi(x)}:\sqrt{L\mu}:e^{-i2\sqrt\pi\widetilde\phi(y)}:|\theta>=N_\theta^2\left(e^{-i\pi\frac{\Delta x}L}\theta_3(\frac{\pi\Delta x}L,e^{-\pi\frac\beta L})\right)^\ast.
$$

The crossed functions computation proceeds as follows

$$<\theta|\sqrt{L\mu}:e^{i2\sqrt\pi\phi(x)}:\sqrt{L\mu}:e^{i2\sqrt\pi\widetilde\phi(y)}:|\theta>=\hskip 4cm$$
\bear\label{crossedsum}
\hskip 1cm&=&N_\theta^2 \sum_{n,n^\prime}e^{-\beta(E_n+E_{n^\prime})/2}<0,0|{U_0^\dagger}^n{\widetilde U_0^\dagger}{}^ne^{i2\pi\frac xL(\Pi_0+\frac 12)+iq+if(c_3+c_3^\dagger)}e^{i2\pi\frac yL(\widetilde\Pi_0-\frac 12)+i\widetilde q+if(\widetilde c_3+\widetilde c_3{}^\dagger)}{U_0}^{n^\prime}{\widetilde U_0}^{n^\prime}|0,0>\nonumber\\
&=&N_\theta^2 \sum_{n,n^\prime}e^{-\beta(E_n+E_{n^\prime})/2}<0,0|e^{i\frac{2\pi}L\left(x(\Pi_0+n+\frac 12)+y(\widetilde \Pi_0-n-\frac 12)\right)}{U_0^\dagger}^n{\widetilde U_0^\dagger}{}^n{U_0}^{n^\prime+1}{\widetilde U_0}^{n^\prime+1}|0,0>\nonumber\\
&=&N_\theta^2\sum_ne^{-\beta(E_n+E_{n+1})/2}e^{i{2\pi}(n+\frac 12)\frac {\Delta x}L}=N_\theta^2 e^{-\frac{\beta\pi}{4L}}\sum_ne^{-\frac\beta L(n+\frac 12)^2+i{2\pi}(n+\frac 12)\frac {\Delta x}L}\nonumber\\
&=&N_\theta^2 e^{-\frac{\beta\pi}{4L}}\theta_2(\frac{\pi\Delta x}L,e^{-\pi\frac\beta L})
\ear

With this results we obtain the equations (\ref{diract}), (\ref{diract2}) and (\ref{diractcrossed}). The Jacobi theta-functions
in the denominators of these relations  are obtained from the computation of the expected values on the vacuum of the thermalized normal modes given in the equations (\ref{mebeta}) and (\ref{mebetatilde}). The theta-functions in the numerators are obtained
from the contributions of the zero modes, computed as expected values of zero-temperature fields on the theta-state.

 It is important to note that the zero mode contribution computation can be interpreted
as a sum over energy states. These energy states are to be associated to the action of the ladder operator $U_0$ on
the zero energy states. Two aspects should, however, be stressed. First, the eigen-energies are obtained as a result of the computation of the expected values of $H_0$ on the vacuum. In the indefinite norm formalism, the ladder operators do not strictly create eigenstates of $H_0$ when acting on the vacuum. Second, the ladder operators were defined in order to
 retrieve the fermion two-point functions from the Mandelstam formula. In this sense this computation is not intrinsically derived from the quantization of the scalar field. For instance,  defining the ladder operators with a diverse factor in the exponential, $U_0\rightarrow U_0^\alpha=e^{i\alpha q}$, would lead to a redefinition of the expected values of the Hamiltonian, and to other two-point functions, which would not furnish the expected bosonization formulas. This arbitrariness, however, is not intrinsic the indefinite norm formalism, but an structural aspect underlining the bosonization process.

\appendix{{\centerline{\bf{Appendix C }}}}
\vspace{0.5cm}

\centerline{\bf Four point functions}

\renewcommand{\theequation}{{C}.\arabic{equation}}\setcounter{equation}{0}

The four-point functions and similarly the N-point functions can be derived from the bosonic representation. Some subtleties deserve to be addressed, however.

For notational economy let us again consider only one of the chiral components and drop the chiral label as in the previous section.
From the fermionic point of view the four-point function is retrieved through the Wick theorem as

\bear\label{fourF}
S^F_{1234}&\equiv&  <0|\psi(x_1)\psi^\dagger(x_2)\psi(x_3)\psi^\dagger(x_4)|0>\nonumber\\
&=&S_{12}S_{34}-S_{14}S_{32}\nonumber\\
&=&\alpha_F\left[\frac{\theta_3(x_{14})\theta_3(x_{32})} {\theta_1(x_{14})\theta_1(x_{32})}    - 
\frac{\theta_3(x_{12})\theta_3(x_{34})} {\theta_1(x_{12})\theta_1(x_{34})} \right]
\ear                    
with the convention that $x_{ij}=(x^\pm_i-x^\pm_j)\frac\pi L$. Also $S_{12}=<0|\psi(x_1)\psi^\dagger(x_2)|0>$.

In the bosonic language, however, we have to compute the expected value of a string of exponentials
of free bosonic fields. In the case of  in infinite space at zero or finite temperature,
and in the case of  of cyclic space fields at zero temperature the expected values can
be reduced to to computation of {\bf vacuum-expected values}. In the case of finite temperature infinite
space fields this is accomplished with the use of thermofield-bosonization as has been shown in \cite{ABR,ABR2}.
In these cases, the four point function is obtained from the bosonic point of view, after the computation of the expected value of a string of four exponential operators, as the exponential of six terms, which can be recast in the form:

\be\label{prop1}
S^B_{1234}=\alpha_B\frac{S_{12}S_{34}S_{14}S_{32}}{S_{13}S_{24}}
\ee

Algebraic properties of the explicit two-point functions  allow to identify

\be\label{fourpart}
S^F_{1234}\propto\frac{S_{12}S_{34}S_{14}S_{32}}{S_{13}S_{24}},
\ee
in all the cases, except when both finite temperature and space periodicity holds. In those cases, the identification follows straightforwardly from the property $
(S_{12})^{-1}=F(x_1)G(x_2)-F(x_2)G(x_1)$, 
which is satisfied by the functions ${x_{12}}$, ${\sinh(x_{12})}$ and ${\sin(x_{12})}$, associated respectively to the cases of zero-T and infinite space, non-zero-T and infinite space and, thirdly, zero-T and periodic space. However in the case of concurrence of both finite temperature and space periodicity the equality in eq. (\ref{prop1}) does not hold. A graphical analysis establishes its failure.

Another difference  in this case is that the N-point functions are not obtained from the computation of vacuum-expected
values, as we have seen. The distinction comes from the computation of the zero mode contribution. In this case we have to compute

\be
S^B_{1234}=S^{B-0}_{1234}\times S^{B-NM}_{1234}= S^{B-0}_{1234}\times\alpha^{NM}_B\frac{S^{NM}_{12}S^{NM}_{34}S^{NM}_{14}S^{NM}_{32}}{S^{NM}_{13}S^{NM}_{24}},
\ee
where the upper-scripts ${S}^{B-0}$ and ${S}^{B-NM}$ refer to zero-modes and normal-modes. The computation of the zero mode contribution to the four-point functions can be defined in analogous way to the two-point one, inserting the
sum of expected values 
\bear
S^{B-0}_{1234}&\propto&\sum<0|{U^\dagger_0}^n\psi^0_1\psi^{0\dagger}_2\psi^0_3\psi^{0\dagger}_4{U_0}^n|0>\nonumber\\
&=&e^{i\pi\frac{x_{12}+x_{34}}{L}}\theta_3(x_{13}+x_{34})
\ear

The final result is

\be\label{fourB}
S^B_{1234}=\alpha_B\frac{\theta_1(x_{13})\theta_1(x_{24})\theta_3(x_{12}+x_{34})\theta_3(0)}{\theta_1(x_{12})\theta_1(x_{34})\theta_1(x_{14})\theta_1(x_{23})}
\ee

The equality, of (\ref{fourF}) and (\ref{fourB}) is indeed a consequence of the known additions formulas for the Jacobi theta functions. Inspecting the relations in \cite{MO} pg. 377, one obtains 

$$2\theta_1(u)\theta_1(v)\theta_3(w)\theta_3(z)=(\theta_0(u^\prime)\theta_0(v^\prime)\theta_2(w^\prime)\theta_2(z^\prime)-{}_0\leftrightarrow {}_2)+({}_0\rightarrow{}_1,{}_2\rightarrow{}_3),
$$
where the primed variables are linear combinations of the non-primed ones,\cite{MO}.
The sum of the expressions with $u=x_{12}$, $v=x_{34}$, $w=x_{14}$ and $z=x_{32}$ with the choice
$u=x_{14}$, $v=x_{23}$, $w=x_{12}$ and $z=x_{34}$, leads to the desired result
$$
\theta_1(x_{24})\theta_1(x_{13})\theta_3(0)\theta_3(x_{12}+x_{34})=\theta_1(x_{12})\theta_1(x_{34})\theta_3(x_{14})\theta_3(x_{32})+\theta_3(x_{12})\theta_3(x_{34})\theta_1(x_{14})\theta_1(x_{23}).$$
From here, it follows the complete four-point function computed with the bosonic representation satisfy the property in eq. (\ref{fourF}) that we wanted to establish.

 We have thus verified the equivalence of the bosonic and fermionic descriptions for the four point function, even with the failure of the property in eq. (\ref{fourpart}).

\end{document}